\documentclass[aps,prb,10pt,twocolumn,superscriptaddress]{revtex4}

\usepackage{graphicx}
\usepackage{amssymb,amsmath}
\usepackage{dcolumn}

\newcommand{\urs}{URu$_2$Si$_2$}
\newcommand{\liii}{$L_\text{III}$}

\begin{document}

\title{Probing $5f$-state configurations in URu$_2$Si$_2$ with 
U $L_\text{III}$-edge resonant x-ray emission spectroscopy}

\author{C.~H. Booth}
\affiliation{Chemical Sciences Division, Lawrence Berkeley National Laboratory, Berkeley, California 94720, USA}
\author{S.~A. Medling}
\affiliation{Chemical Sciences Division, Lawrence Berkeley National Laboratory, Berkeley, California 94720, USA}
\affiliation{Electronic Materials Engineering, Australian National University, Canberra, 2602, Australia}
\author{J.~G. Tobin}
\altaffiliation{Present address: Departments of Chemistry and Physics, 
University of Wisconsin-Oshkosh, WI 54901, USA}
\affiliation{Materials Science Division, Lawrence Livermore National Laboratory, Livermore, California 94550, USA}
\author{R.~E. Baumbach}
\altaffiliation{Present address: National High Magnetic Field Laboratory, Tallahassee, FL 32310, USA}
\affiliation{Materials Physics and Applications Division, Los Alamos National Laboratory, Los Alamos, New Mexico 87545, USA}
\author{E.~D. Bauer}
\affiliation{Materials Physics and Applications Division, Los Alamos National Laboratory, Los Alamos, New Mexico 87545, USA}
\author{D. Sokaras}
\affiliation{Stanford Synchrotron Radiation Lightsource, SLAC National Accelerator Laboratory, Menlo Park, California 94025, USA}
\author{D. Nordlund}
\affiliation{Stanford Synchrotron Radiation Lightsource, SLAC National Accelerator Laboratory, Menlo Park, California 94025, USA}
\author{T.-C. Weng}
\altaffiliation{Present address: 
Center for High Pressure Science \& Technology Advanced Research,
1690 Cailun Rd., Bldg \#6, Pudong, Shanghai 201203, P. R. China}
\affiliation{Stanford Synchrotron Radiation Lightsource, SLAC National Accelerator Laboratory, Menlo Park, California 94025, USA}

\date{To be published in Physical Review B}

\begin{abstract}
Resonant x-ray emission spectroscopy (RXES) was employed at the U \liii\ 
absorption edge and the $L_{\alpha1}$ emission line to explore the
$5f$ occupancy, $n_f$, and the degree of $5f$ orbital delocalization in the
hidden order compound \urs. By comparing to suitable reference materials
such as UF$_4$, UCd$_{11}$, and $\alpha$-U, we conclude that the $5f$ orbital in
\urs\ is at least partially delocalized with $n_f = 2.87 \pm 0.08$, and
does not change with temperature down to 10 K within the estimated error. These
results place further constraints on theoretical explanations of the
hidden order, especially those requiring a localized $f^2$ ground state.
\end{abstract}

\pacs{71.27.+a, 78.70.Dm, 78.70.En, 75.20.Hr}

\maketitle

\section{Introduction}
\label{Intro}

It has often been said that the most interesting physics occurs when
competing interactions are of nearly the same magnitude. Such a situation is 
surely occurring at URu$_2$Si$_2$'s so-called ``hidden-order transition'' (HO), 
which is a second-order phase transition at 17.5 K with a large 
($0.2 R \log2$, where $R$ is the universal gas constant) change in entropy 
that has nevertheless so far defied
attempts to identify its order parameter.  \cite{Palstra85,Maple86,Broholm87}
Despite this conundrum being identified in the late 1980s and the 
fact that the HO transition presages a 1.5 K
superconducting transition, the identification of the HO
order parameter remains elusive, although progress 
has been steady since that time and \urs\ 
remains an important research subject today.\cite{URS_Special_Issue14}
Much of the recent work has been spurred by new innovative theories and 
concomitant improvements to experiments made possible both by better capabilities and
single crystals. An important dividing line between the different
theories of HO focuses on the nature of the $5f$ orbital, specifically, the
$5f$-orbital occupancy, $n_f$, and the degree of itinerancy.\cite{Mydosh14}
Various spectroscopic measurements of these quantities have been performed, 
without a clear consensus.  In an effort to clarify the role of $n_f$ and
$5f$ localization, the work described below provides measures of both 
$n_f$ and the degree of $5f$-orbital localization using resonant x-ray emission 
spectroscopy (RXES) at the U \liii\ absorption edge and 
U $L_{\alpha1}$ emission line.

The history of theoretical work describing HO in \urs\ is 
vast.\cite{Mydosh11,Mydosh14} For this short introduction, we only focus on 
some specific aspects. Some of the earliest theories relied on the existence
of a localized $f^2$, $J=4$ configuration to generate certain crystalline
electric field (CEF) symmetries. Although CEF signatures have never been
definitively observed, some recent innovative work once more depends on their 
existence,\cite{Chandra15,Chandra13,Toth11,Su11,Harima10,Haule09} while other 
work focuses on an itinerant model of the $5f$ electrons starting from a 
partially occupied $f^3$ orbital.\cite{Oppeneer10,Oppeneer11} The DFT+DMFT
calculations may form an interesting intermediate starting point, assigning
the CEF states to the $j=5/2$ shell and itinerant states to the $j=7/2$ shell.\cite{Haule09,Kung15,Kung15SI}

Experimental investigations are similarly divided in their interpretations.
For instance, neutron scattering results show that 
a spin excitation gap 
can explain the change in the specific heat at 17.5~K, but
is not consistent with localized physics.\cite{Wiebe07}
Likewise, recent NMR experiments looking at Knight shift anomalies are
modeled such that the HO emerges directly from a Kondo liquid state, and is 
thus not associated with localized moments.\cite{Shirer12}
In addition, neutron scattering has not definitively observed any CEF states.\cite{Butch15}
On the other hand, thermal conductivity measurements indicate a transition from itinerant to 
localized behavior in the HO state.\cite{Behnia05}
Other indications of at least a partially localized $f^2$ configuration 
exist, together with indications of dynamical CEF 
excitations.\cite{Wray15} 
In addition, recent experiments highlight the possible importance of symmetry
changes. For instance,
cyclotron resonance measurements show an anomalous splitting of the sharpest resonance line
under in-plane magnetic field rotation, likely caused by the fourfold rotation
symmetry of the tetragonal lattice being broken by domain formation, and
consistent with the suggestion that there is a nematic Fermi liquid state
(where itinerant electrons have unidirectional correlations).\cite{Tonegawa12}
This result is supported by high-resolution synchrotron x-ray diffraction
results.\cite{Tonegawa14}
Other measurements indicating possible tetragonal symmetry breaking in the 
HO state include a recent measurement of a long-lived low energy 
excited-state chirality density wave 
with $A_{2g}$ symmetry from Raman spectroscopy,\cite{Kung15} 
consistent with inelastic neutron scattering 
anisotropy results.\cite{Bourdarot03}
However, this
conclusion remains controversial; for instance, recent inelastic neutron 
results show no indications of reduced spatial symmetry,\cite{Butch15}
raising the possibility that such symmetry breaking only happens in the 
smaller samples less suitable for neutron experiments.
Moreover, comparisons of the DC magnetic susceptibility $\chi(T)$ of a system 
thought to possess a tetravalent singlet crystal field ground state similar to 
that proposed for \urs\ show little resemblance.\cite{Grauel92} 

Ultimately, determining specific and quantitative details about $n_f$ and
$5f$ itinerancy require spectroscopic measurements.  Photoemission (both 
angle-integrated and angle-resolved) results generally favor delocalized
$5f$ states and paint a very interesting picture of the details of the
Fermi surface.\cite{Durakiewicz14}
In particular, the larger features in the band structure and Fermi surface of \urs\ measured 
by soft x-ray photoemission in the paramagnetic (PM) state above the HO 
transition are well 
explained by treating all of the U $5f$ electrons as itinerant with
$n_f \approx 2.6$.\cite{Kawasaki11}
Furthermore, these photoemission experiments
indicate a large electron-like sheet around the $\Gamma$ point, with smaller
hole-like structures forming around the $Z$ point.\cite{Kawasaki11} It is
important to note, however, that not all features in photoemission are
well described by LDA calculations. For instance, some 
indications of an $f^2$ contribution have also been observed
in core-level and valence-band photoemission that are otherwise indicating
a close to $f^3$ ground state.\cite{Fujimori12}
In addition,
changes in the HO phase include a Fermi surface 
restructuring\cite{Santander-Syro09} involving folding along $Q_0 = (0, 0, 1)$
and gapping along the $(1, 1, 0)$ directions.\cite{Meng13}

Although the photoemission experiments have provided valuable insight into the
electronic structure of \urs, they are
limited by surface-sensitivity concerns and by having to compare to 
calculations in order to determine a value for $n_f$.\cite{Durakiewicz14} 
One technique that is less sensitive to the exact electronic details is 
electron energy-loss spectroscopy (EELS). EELS experiments suggest 
that \urs\ has $5f$ states which are more localized than $\alpha$-U (even at 
room temperature), but still not completely localized, with a $5f$ electron 
count $n_f =  2.7 \pm 0.1$ suggesting a mixed valence ground state and/or some 
$5f$-electron itinerancy.\cite{Jeffries10} However, other interpretations 
question the usefulness of EELS in this respect in uranium 
compounds,\cite{Tobin15} implying that further support is needed.
In addition, the degree of surface sensitivity in this electron 
spectroscopy technique remains a concern.

Photon-in/photon-out techniques are inherently less surface sensitive,
although a soft x-ray experiment of this type is much more 
surface sensitive than a hard x-ray experiment.
A recent $O$ edge x-ray absorption spectroscopy (XAS) measurement, combined 
with resonant inelastic scattering (RIXS) and including polarization 
dependence,\cite{Wray15}
concludes that only features 
derived from an $f^2$ ($J=4$) ground state are clearly observed; however, it is 
noted that $f^3$ features could be obscured by itinerancy, which has not yet 
been considered with experimental model compounds. The measurements observed 
only short-lived ($\Gamma \gtrsim 0.1$ eV) CEF modes, and found that these could
account for most or all of the CEF excitation intensity expected in an $f^2$ 
multiplet picture. 
Taken together, although features associated with 
the $f^2$ atomic multiplet ground state ($J=4$) are clearly observed, the total 
$5f$ occupancy is not determined by these data. Likewise, the large inverse 
lifetime of CEF excitations (0.1 eV $\gg 1/k_BT$) implies that electronic 
itinerancy cannot be dismissed as a perturbative factor.

An independent, truly bulk-sensitive method for determining $n_f$ is clearly highly desirable.
U \liii\ RXES should be able to provide such an independent measure of $n_f$, 
while potentially also drawing a distinction between a Kondo-like, mixed
valence mechanism and a $5f$-band interpretation, but there are challenges. 
The technique involves
measuring the U $L_{\alpha1}$ x-ray emission as a function of energy using a
high resolution spectrometer while sweeping the incident x-ray energy just
above and below the U \liii\ absorption edge. The average depth of an emitting
photon above the photoelectron threshold energy is about 
1.9~$\mu$m,\cite{Troger92} and is deeper below the threshold where much of the
data and analysis occur. The resulting spectral broadening
is dominated by the final-state 3$d_{5/2}$ core hole lifetime, and hence
provides a higher resolution measure of the unoccupied 6$d$ states near
the Fermi energy, $E_F$, than a conventional \liii-edge x-ray absorption near-edge
structure (XANES) experiment, which is dominated by the shorter-lived $2p_{3/2}$
core hole. Either RXES or XANES experiments can potentially
differentiate between a localized mixed valence state and a simple 
partially-filled band if the Coulomb interaction between the core hole and
the $f$-electrons is strong enough to break the mixed valence state into its
configurations with different numbers of $f$-electrons.\cite{Kohn82a} In a typical Yb 
intermetallic, for instance, the Coulomb interaction splits the $4f$ state into
$f^{13}$ and $f^{14}$ configurations, which screen the outgoing photoelectron
differently, resulting in two distinct features in the 2$p$-5$d$ absorption
spectrum that are about 10 eV apart.\cite{Sarrao96,Booth10} Such splitting in 
uranium should be approximately the same as observed between valence states, 
which is on the order of $\approx 5$~eV based on studies of various oxide 
materials.\cite{Allen97,Conradson04} On the other hand, if the $5f$-electrons 
are more delocalized, an overall shift of the main absorption feature may occur 
instead of split features, since the Coulomb interaction may then be of 
insufficient strength. These complications need to be considered when analyzing 
either XANES or RXES U \liii\ spectra.

A further complication can occur in the presence of strong ligand fields, where
splitting can occur between $t_{2g}$ and $e_g$ states in the $d$ manifold. If 
this splitting approaches the $\approx 5$~eV expected between valence states in U, then
deconvolving ligand-field splitting and intermediate valence effects 
may not be possible. However, one expects that such ligand-field splitting 
should be relatively small in an intermetallic compound like
\urs\ compared to a more $\pi$ bonded system like UO$_2$. 

In consideration of these effects, the rest of this paper is organized as 
follows: After a description of the Experimental Details, RXES results
from UO$_2$ and UF$_4$ will be compared to those from UCd$_{11}$ as examples of
standard materials exhibiting various degrees of localized $5f$ behavior, 
ligand-field splitting in the $d$ manifold, and both $f^2$ and $f^3$ 
spectroscopic features. Subsequently, results from \urs\ will be presented and
considered in light of potential localized/delocalized behavior and 
ligand-field splitting.

\section{Experimental details and methods}
\label{Exp}

A single crystal of \urs\ was grown by the Czochralski
technique and subsequently electro-refined. Two samples were cleaved from this
crystal. Such cleaves routinely yielded high-purity crystals with 
residual resistivity ratios RRR = $\rho$(300 K)/$\rho$(0 K) between 200-400,
where $\rho$(0) was obtained from a power law fit to the electrical resistivity 
of the form $\rho(T) = \rho_0 + AT^n$ at low temperatures. The RRRs for these
specific cleaved samples for the RXES experiments were not measured. While 
each sample 
was chosen for the spectroscopic measurements to have
an optically flat portion for easy sample alignment, after preliminary
measurements, a single sample was chosen for the measurements presented
here.

RXES data were collected during two experimental runs about one year apart
at the Stanford Synchrotron Radiation Lightsource (SSRL) wiggler beamline 
6-2 using a LN$_2$-cooled Si
(311) double monochromator calibrated so that the inflection point of the
Zr $K$-edge absorption from a Zr reference foil was at 17998.0 eV.  
The emission was measured using a
seven-crystal Ge(777) Johann-type x-ray emission spectrometer,\cite{Sokaras13}
 at an emission
energy, $E_e$, of approximately 13.6 keV, corresponding to the U $L_{\alpha 1}$ 
emission. The emission spectrometer energy was calibrated using the direct 
scatter
from a polycarbonate film with the incident energy, $E_i$, set to the first 
inflection point of the absorption at the Au $L_2$ edge from a Au reference
foil (13734 eV). The total emission energy resolution (including the incident 
beam) was measured to be 1.4 eV. 

At these energies, the information depth of the x-rays is greater than 
1.9~$\mu$m,\cite{Troger92} so these measurements are truly bulk sensitive.
The sample was visibly shiny for both experimental runs, and no particular care 
was taken to avoid surface oxidation. 

The sample of \urs\ was placed
with its surface normal at a 45$^\circ$ angle with respect to the incoming 
beam.  Data were collected at 10, 15, 20, 22, 50, 90, and 300 K using a
LHe-flow cryostat.  
Owing to the relative thickness of the sample, a self-absorption correction
was applied,\cite{Booth14} as well as a dead-time correction.

The RXES emission intensity data, $I_e$, are fitted with previously published 
methods\cite{Booth14}, using the Kramers-Heisenberg equation of the form:
\begin{align}
I_e(E_i,E_t)=\int d\epsilon\  \eta(\epsilon)\frac{A}{(E_{gi}-\epsilon+E_i)^2+\Gamma_i^2/4}\times \notag \\
\frac{\Gamma_f/(2\pi)}{(E_{if}-\epsilon+E_t)^2+\Gamma_f^2/4}.
\label{Ie_eq}
\end{align}
Here, $E_t = E_i - E_e$ is
the energy transferred to the sample in the final state, 
$E_{gi}$ is an energy scale corresponding to 
the energy difference between the ground and intermediate state, $E_{if}$ is
another energy scale corresponding to the energy difference between the
intermediate and the final state, $\Gamma_i$ is the lineshape broadening due to
the finite lifetime of the intermediate state core hole (here, the $2p_{3/2}$
core hole), and $\Gamma_f$ is similarly due to the finite lifetime of the
final state core hole (here, the $3d_{5/2}$ core hole). For a more complete 
discussion of Eq. \ref{Ie_eq}, please see Refs. \onlinecite{Rueff10,Booth14}. 
In the fits
described below, we have chosen to fix $\Gamma_i$ and $\Gamma_f$ to their 
nominal values\cite{Keski74} of 8.104 eV and 3.874 eV, respectively, although 
allowing these parameters to float generally gives results close to these 
values and does not significantly change the results described below. In these 
experiments,
the ground state includes $2p_{3/2} 3d_{5/2} 6\bar{d}$ electrons, the 
intermediate state has $2\bar{p}_{3/2} 3d_{5/2} 6{d}$ electrons, and the final
state has $2p_{3/2} 3\bar{d}_{5/2} 6d$ electrons, where the bar indicate a hole.
Eq.~\ref{Ie_eq} is simplified assuming the transition matrix elements $T_1$ and $T_2$ 
in $A\propto\langle f | T_2 | i \rangle^2\langle i | T_1 | g \rangle^2$
have no off-diagonal terms. 

The most important aspect of the fitting is the
choice of the local unoccupied density of states $\eta(\epsilon)$. As described 
in Ref. \onlinecite{Booth14}, we allow for three different possible $5f$
configurations within the ground state:
\begin{equation}
| f \rangle = c_2 | f^2 \rangle + c_3 | f^3 \rangle + c_4 | f^4 \rangle,
\label{config_eq}
\end{equation}
where $c_i^2$ give the probability of finding the system in any one
configuration $f^i$. The presence of the core hole in both the intermediate
and the final state will interact differently with each configuration, and if
this Coulomb interaction is large enough, these states will split.\cite{Kohn82a}
This splitting is reflected in the empty 6$d$ states. As 
before,\cite{Booth14,Tobin15,Soderlind16} we parametrize $\eta(\epsilon)$ with a
combination of a so-called ``peak'' Gaussian (each constrained to the same width
$\sigma_p$) to represent the excitations into the comparatively discrete empty 
$6d$ states and a broadened step function (same $\sigma_p$ and the height of 
the peak Gaussian defined relative to the 
step height defined to be the $p/s$ ratio) to represent the continuum of 
unoccupied states. Each potential configuration is then represented by this 
combination of a Gaussian and the step function. More details with regard to the
specific fits are provided below.

\section{Results}
\label{Results}

\subsection{Calculations and measurements on UF$_4$ and UO$_2$}

In order to consider the $5f$-orbital occupancy and localization features of
\urs, comparisons to standard materials are essential. In this case, the 
standard materials would ideally be ones with strongly localized $5f$-orbitals
in the $f^2$ (tetravalent uranium) and $f^3$ (trivalent uranium) configurations.
From our previous work,\cite{Booth12,Soderlind16} we identified UCd$_{11}$ as 
possessing strongly localized $5f$ electrons and $n_f = 2.86 \pm 0.08$, which is 
sufficiently close to $f^3$ to act as a good standard.\cite{Soderlind16} 
Unfortunately, although there are only a few intermetallics thought to possess 
a localized $f^2$ configuration, we have not
succeeded in obtaining data on sufficiently localized intermetallic samples of 
this type. Instead, we can rely on data from UF$_4$ as an unquestionably 
localized $f^2$ material.\cite{Tobin15}

\begin{figure}
\includegraphics[width=3.2in]{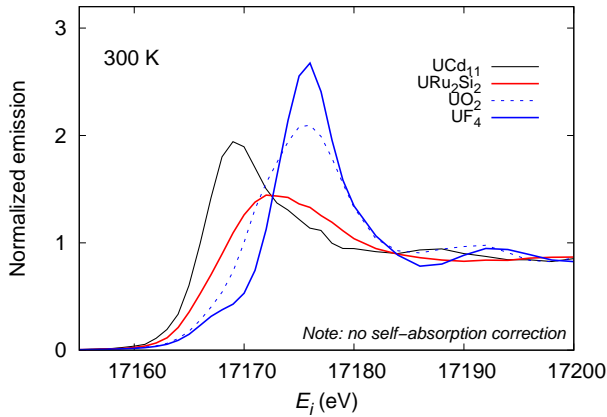}
\caption{PFY measurements of UCd$_{11}$ (black), \urs\ (red), UO$_2$ (blue 
dotted), and UF$_4$ (blue) each collected at 300 K. Note that a self-absorption 
correction is not applied to these data to better place them on the same scale 
and to accentuate the feature in the UF$_4$ spectrum on the lower shoulder of 
the main edge near 17167 eV.
}
\label{pfy_fig}
\end{figure}

High-resolution partial fluorescence yield (PFY) data are shown in Fig. \ref{pfy_fig} for all measured 
samples. As indicated, the self-absorption correction was not applied to the 
data in this figure as a convenient way to accentuate the clear shoulder peak 
at about 17167 eV in the UF$_4$ spectrum. As is clearly seen, the so-called
``white line'' (WL) peak in UCd$_{11}$ is shifted by 7-8 eV relative to that of 
UF$_4$. The UO$_2$ and \urs\ spectra are clearly broader, and the WL peak
energy of \urs\ is between that of UCd$_{11}$ and UF$_4$.

There are two features of the standards spectra that can be elucidated with
cluster calculations, namely the broadening of the UO$_2$ spectrum and the
shoulder feature in the UF$_4$ spectrum. Starting with the UO$_2$ spectrum, we
note that in previous work, we have used UO$_2$ as a localized $f^2$ standard, 
but have 
recently found it to be a problematic example. UO$_2$ is considered to be a 
correlated-electron material and a Mott-Hubbard insulator,\cite{Dudarev97,Yu11} and as such, it
may have a $5f$ occupancy that deviates from two and even have some direct 
$5f$-band involvement at the Fermi energy. More importantly, the ordered cubic 
symmetry and octahedral coordination of the U-O nearest neighbors generates
a substantial crystal field splitting of the unoccupied $d$ states, a
situation that is reduced in the more complex monoclinic structure of UF$_4$.

\begin{figure}
\includegraphics[width=3.2in]{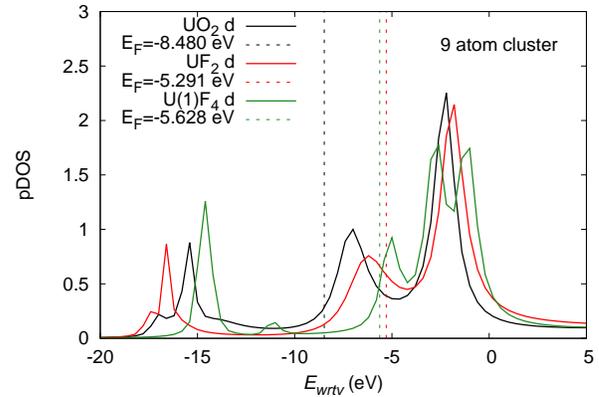}
\caption{Local density of states from FEFF LDA approximation for a 9 atom cluster of UO$_2$ (black), UF$_2$ (based on the UO$_2$ structure, red), and the U(1)
site in UF$_4$ (green). $E_F$ for each calculation is shown as
a vertical dotted line.
}
\label{pDOS_fig}
\end{figure}

\begin{figure}
\includegraphics[width=3.2in]{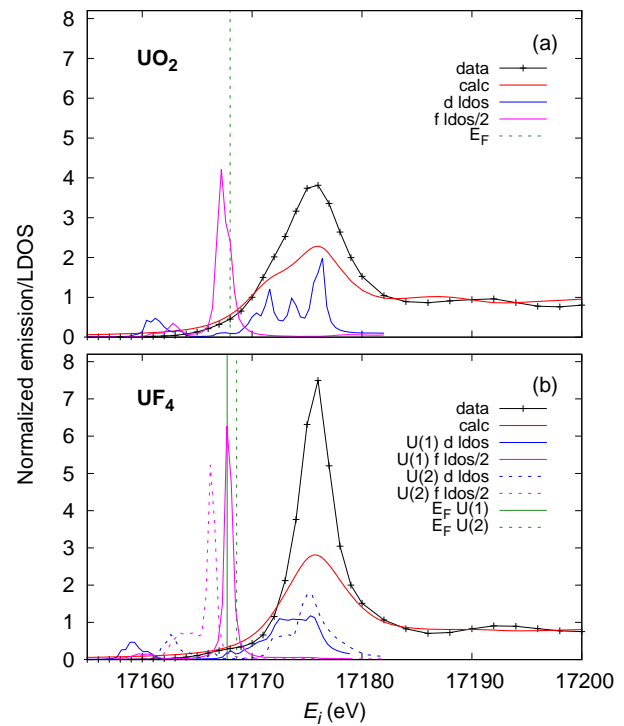}
\caption{PFY data shown together with a narrowed XANES and the corresponding
local density of states calculations from FEFF LDA approximation for a 9 atom 
cluster of (a) UO$_2$ and (b) UF$_4$. Calculations are shifted 
by (a) 17177.0 eV and (b) 17173.9 eV relative to the vacuum energy.
}
\label{pfycalc_fig}
\end{figure}

This situation is illustrated by the results in Fig. \ref{pDOS_fig} of a 9 atom 
cluster calculation of the local $d$ density of states using 
FEFF 9.6.4.\cite{FEFF9} In this simplified calculation, we use default FEFF
behavior, which includes not allowing for charge transfer out of the
$5f$ orbitals. Three curves are shown. In each calculation, only
the first shell of 8 oxygen or fluorine atoms are included along with
the absorbing uranium atom. The small cluster size was chosen to emphasize the
short-range, ligand-field nature of the $e_g$ and $t_{2g}$ features. The
UO$_2$ calculation uses the nominal fluorite structure\cite{Wyckoff} and the
UF$_4$ calculation uses the nominal monoclinic structure.\cite{Larson64} To
demonstrate the role of the fluorine atom as opposed to the difference
in crystal structure, we also show a calculation on ``UF$_2$", which is really
the same calculation on the same structure as the UO$_2$ calculation, except
all the oxygen atoms are replaced by fluorine.

A number of features are important to discuss with regard to how UO$_2$ and/or
UF$_4$ make a suitable localized $f^2$ U \liii\ absorption standard. In all
three calculations, there is an $e_g$ state, moving from about -7 eV in UO$_2$
(with respect to the vacuum energy) to about -5 eV in UF$_4$. The
$t_{2g}$ state is at a somewhat higher energy, all centered at about -2~eV, 
with the UF$_4$ calculation showing a $\approx$ 1.5 eV split. These calculations
therefore demonstrate that the $e_g$/$t_{2g}$ ligand field splitting is
reduced from UO$_2$ as one moves to the more ionic/less covalent ``UF$_2$'' 
compound and then further to the less symmetric UF$_4$ compound.

These differences can be seen in the resulting absorption calculations
shown in Fig. \ref{pfycalc_fig}, where we now show the calculations on the
``real'' UO$_2$ and UF$_4$ structures for comparison to actual data 
(which now includes the self-absorption correction). 
These calculations include all atoms within a
6.58 \AA\ radius of the central absorption uranium, and take into account
the two uranium sites in the UF$_4$ structure, as noted. The spectra are 
calculated as \liii-edge absorption spectra but narrowed by 4.2 eV (FWHM) to
account for the limiting factor of the $3d_{5/2}$ core hole instead of the 
$2p_{3/2}$ core hole. In addition, 
charge transfer out of the $5f$ orbital is allowed (the UNFREEZEF card is
employed). In order to obviate the $\approx$ 0.05 \% threshold energy errors 
in FEFF calculations, the calculations are shifted 
by the amounts indicated in the figure caption
so that the WL energies
agree with the data.

While one can clearly see the effect of the larger ligand
field splitting in the UO$_2$ calculation and the data compared to those of 
UF$_4$, it is clear that FEFF overestimates the size of this splitting in
each case. This overestimate is particularly clear in the UF$_4$ calculation,
which shows more weight than the experimental data near 17170 eV, in a region
of the spectrum between the shoulder feature at 17167 eV and the main edge.

A fascinating feature of the calculation on UF$_4$ is the difference in 
$E_F$ between the U(1) and U(2) sites. While bearing in mind that
potential errors exist in the determination of $E_F$, the FEFF
calculations show a distinct energy shift in the $5f$ density of states 
between the U(1) site and the U(2) site (there are double the number of
U(2) sites in the UF$_4$ lattice structure). This shift places the Fermi
level within the U(1) density of states, a situation that also occurs, although 
to a lesser degree, in the UO$_2$ calculation. The significance of this shift 
is that, according to these calculations, unoccupied spectral weight occurs in 
the $5f$ band at or just above $E_F$, which is accessible to the
photoelectron excited from the 2$p_{3/2}$ shell either from a dipole
excitation through hybridization with the $d$ orbitals or directly through a 
quadrupole transition, as previously considered for UO$_2$.\cite{Vitova10}
The position
and size of this feature are in very good agreement with the FEFF calculation 
as shown, which does not include any quadrupole term in the excitation.
Including such a term vastly over estimates the size of the feature, possibly
due to the $5f$ weight at the Fermi energy. We therefore tentatively conclude 
that this feature is primarily due to $f/d$ hybridization, consistent with 
several photoemission studies.\cite{Durakiewicz14}

An interesting issue in these calculations is that the calculated $E_F$ is 
about 3 eV lower in UO$_2$ than the other calculations. This difference exists
in both the small cluster and the 6.58 \AA\ radius calculations.
Fermi energy shifts are a common problem in FEFF, but this particular shift 
may be a reflection of the correlated electron nature
of UO$_2$, a quality that FEFF cannot capture. $E_F$ directly affects 
the photoelectron threshold energy, yet no shift is observed in the 
experimental data
in the white-line position between UO$_2$ and UF$_4$ (Fig. \ref{pfy_fig}).
It should therefore be noted that the
absorption calculation in Fig. \ref{pfy_fig}(a) is with respect to $E_F$,
rather than with respect to vacuum, in order to make a direct comparison
between the calculations and between the calculations and experiment.

The implication of these data and calculations for the purposes of this study
is that the ligand field splitting is a complicating factor in the UO$_2$
spectra when using such spectra to model $f^2$ behavior in metals where
such splitting will be reduced. In contrast, UF$_4$ appears to be a much better,
and less covalent, model for such comparisons. In fact, the ligand field
splitting is even less clear in larger cluster calculations of UF$_4$,
creating an even sharper absorption white line. Therefore,
the combination of more ionic bonding (through the replacement of 
oxygen with fluorine) and reduced symmetry (which further reduces the ligand
field splitting) allows UF$_4$ to be used as a close-to-ideal localized $f^2$ 
absorption standard material.

\subsection{Comparisons between standard materials and URu$_2$Si$_2$}

\begin{figure}
\includegraphics[width=3.2in]{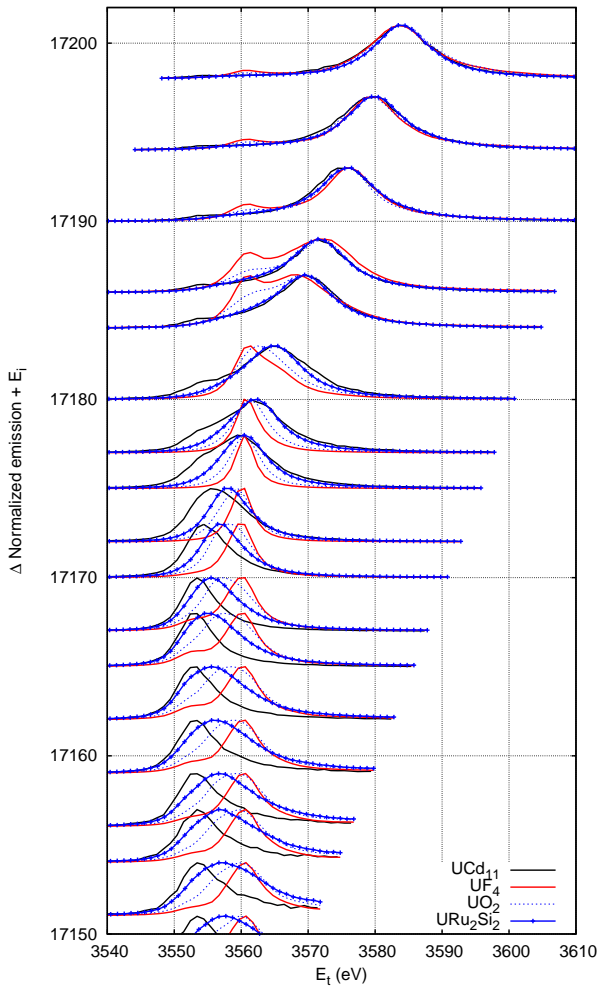}
\caption{Stacking plot for the UCd$_{11}$, UO$_2$, and UF$_4$ standards together
with data from \urs, all collected at 300~K. Each constant-$E_i$ scan is
normalized to maximum emission flux so that data near resonance (very high flux)
can be placed on the same scale as, and thus more easily compared with, data well
below resonance. Note that this normalization scheme obscures the fact that
the unnormalized emission flux below threshold ($\sim 17170$~eV for all
these samples) is significantly lower than the emission flux above threshold.
}
\label{stack_fig}
\end{figure}
\begin{figure}
\includegraphics[width=3.2in]{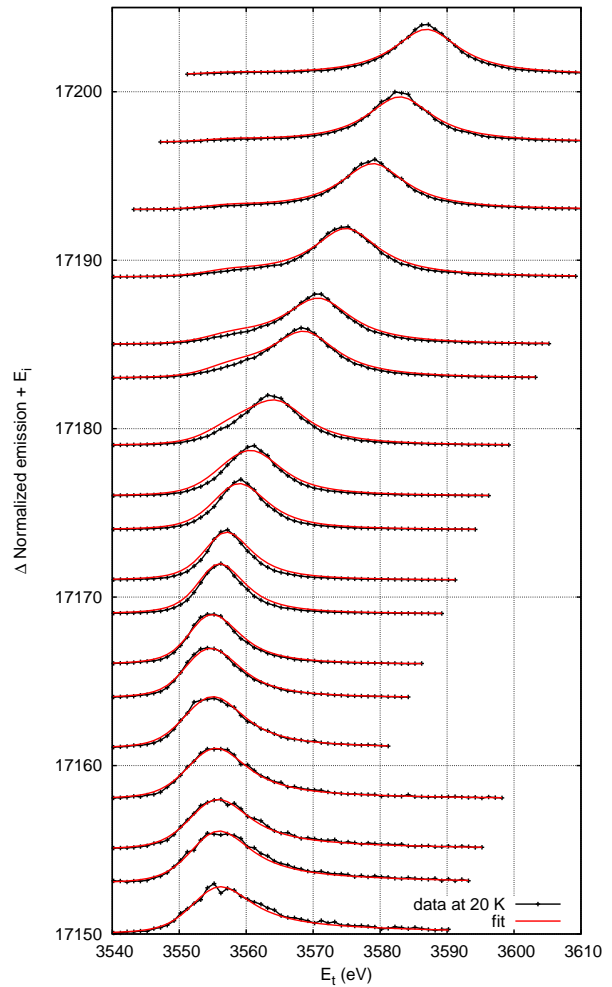}
\caption{RXES data and fit using model 2 at 20 K for \urs.
}
\label{U122fit_fig}
\end{figure}

RXES data at 300 K for the standard materials and \urs\ are shown in 
Fig.~\ref{stack_fig}. We begin with a discussion of the features in the 
various data sets before describing the fit results below. 
First, comparing the different standard materials, differences
are most easily observed and interpreted well below the \liii\ threshold
energy. In this method of presenting the data, the data below
threshold are toward the bottom of the plot and the features in such data
are at a relatively fixed $E_t$. It is clear from these data that the
UF$_4$ and UCd$_{11}$ spectra are each relatively sharp in character, while
separated by about 7 eV, consistent with, but even larger than, the 
$\approx 5$~eV shift expected for a  1 e$^-$ difference
in their $5f$ shells (Sec. \ref{Intro}). The effect of the ligand field splitting in UO$_2$
is pronounced in these data, with a significantly broader spectrum below
threshold. In addition to this broadening, there is a small positional shift of the XES peak (more easily observed at low $E_i$, which is also rationalized by the ligand field splitting. 

The \urs\ data fall between the limits defined by the UCd$_{11}$ and the 
UF$_4$ data, with a significant amount of spectral weight at both extremes.
The lower-energy weight is even more clearly observed as it becomes resonantly 
enhanced near $E_i \approx$ 17166 eV.
It is interesting to compare these \urs\ results to those from UO$_2$, since 
the energy shift toward UCd$_{11}$ is substantially larger and the
spectra are significantly broader compared to those from 
UO$_2$.
Since no large ligand field splitting is expected in the $d$ manifold in
\urs, it seems very unlikely that it could be larger in \urs\ than
in UO$_2$. 

Given the magnitude of the negative energy shift of the \urs\ 
spectra relative to UF$_4$ (and UO$_2$) is too large to be explained by ligand
field splitting, the sign of this shift is significant: Since the XES peak
position of UF$_4$ is determined by a localized $f^2$ configuration, the 
negative comparative shift of the \urs\ spectra indicates a more fully screened
core hole, which indicates more occupied $5f$ weight (not less), that is, an 
$f^3$ component to the wavefunction.  
These simple comparisons therefore yield one of the main conclusions of the
present study: a significant, if not dominant, $f^3$ component to the
\urs\ wavefunction exists.

In addition, the enhanced width of the \urs\ resonance may suggest an 
intermediate
occupancy of the $5f$ orbital, either due to a partially-filled metallic $5f$ 
band, a Kondo-driven intermediate valence effect, or a mixture of both.
Unfortunately, unlike data from $\delta$-Pu\cite{Booth12}, or even
$\alpha$-U,\cite{Soderlind16} there are no spectra collected at any of the 
$E_i$ considered here that show visible indications of multiple contributions to
the main XES peaks indicative of mixed valence. For a more 
quantitative consideration, we turn to the results from the detailed fits.

\begin{table*}
\caption{RXES fit results for data collected at 20 K for \urs\ and for
previously published work\cite{Soderlind16} on UCd$_{11}$ and UF$_4$ as 
examples of localized $f^3$ and $f^2$ compounds, UO$_2$ as an example of a 
system with significant ligand field splitting, and $\alpha$-U as an example of 
a metal with
itinerant $5f$ electrons. $E_{gi}$ and $E_{if}$ are defined in these fits to
correspond to the $f^2$ resonance position. The energy separation between each 
of the $f^2$, $f^3$, and $f^4$ resonances is set to $\Delta E = 7.2$ eV, as
determined from the fits to the UCd$_{11}$ data and the UF$_4$ data. See
Sec. \ref{Exp} for a further description of the fit parameters. Reported
errors assume normally distributed errors and are obtained from the covariance
matrix. Systematic errors are likely
larger for the configuration fractions, and we estimate $\pm 5$\% as a 
reasonable error estimate for both the configuration fractions and $n_f$.
Note that average result of data collected at
difference temperatures and including systematic error gives 
$n_f = 2.87 \pm 0.08$ (see Fig. \ref{nf_fig}). 
}
\begin{ruledtabular}
\begin{tabular}{lcccccccc}
Compound & $E_{gi}$(eV) & $E_{if}$(eV) & $\sigma_p$(eV) & $p/s$ &  $f^2$ (\%) & $f^3$ (\%) & $f^4$ (\%) & $n_f$(e$^-$)\footnote[1]{As with other parameters, reported errors assume normally distributed errors and these fits are for data at only one temperature. Systematic errors on $n_f$ are thought to approach 0.08. See Fig. \ref{nf_fig}.} \\
\colrule
UCd$_{11}$ & 17175.0(1) & 3560.0(1) & 1.9(1) & 3.3(1) &  13.5(7) & 86(1) & 0.1(5) &2.86(1) \\
UF$_{4}$ & 17174.3(1) & 3560.3(1) &  1.2(4) & 9 &  100(3) & 0(2) & 0(1) &2.00(3) \\
UO$_{2}$ & 17173.9(1) & 3559.8(1) & 1.9(4) & 5.6(2) &  95(2) & 5(1) & 0(1) &2.05(2) \\
$\alpha$-U & 17176.9(1) & 3562.5(1) & 2.9(4) & 1.5(1) &  46(2) & 54(3) & 0(2) &2.54(2) \\
\\
\urs\ Model \#1 & 17178.6(1) & 3564.6(1) & 3.4(1) & 1.7(1) & 0 &  100 & 0 & 3 \\
\urs\ Model \#2 & 17178.1(2) & 3564.0(2) & 3.1(4) & 2.0(1) &  11(2) & 89(2)& 0.0(5) &2.89(2) \\
\end{tabular}
\end{ruledtabular}
\label{table}
\end{table*}

Two fit models were considered: Model \#1 allows only a single configuration, while
Model \#2 allows for up to three configurations to exist.
The fit results are summarized in Table \ref{table}, together with previous
results on the standard materials and $\alpha$-U. \cite{Soderlind16} 
The fits are both of high quality and the results from each model are not
easily discernible. The somewhat higher-quality fit uses Model \#2, which is
displayed in Fig. \ref{U122fit_fig}. While this model is significantly
better than Model \#1 in a statistical sense, the fits are not visibly very 
different, and systematic errors (especially due to the line shape)
remain the main contribution to the 
quality-of-fit parameter (proportional to a statistical-$\chi^2$).
We therefore do not make a 
judgment here as to which fit model is more appropriate; in 
Sec. \ref{Discussion} we discuss some reasons for favoring Model \#2, although
the results aren't very different.  Fit methods are described in 
Ref. \onlinecite{Booth14}. In particular, the fits utilize a parametrized
$\eta(\epsilon)$ where the contribution to each resonance includes a Gaussian 
peak and an arctan-like function (an integrated pseudo-Voigt).  The energy 
scale $E_{gi}$ is defined here as the excitation energy from the 2$p_{3/2}$
shell into the unoccupied states associated with the $f^2$ configuration. The
$E_{if}$ energy scale is defined similarly.  
The Model \#2 fits
assume an intermediate valence model where the Coulomb interaction is 
sufficient to split the potential $f^2$, $f^3$, and $f^4$ configurations,
using a fixed energy separation of 7.2 eV as determined 
previously.\cite{Soderlind16} Peak assignments (including in Model \#1) are 
assigned relative to the main $f^2$ peak in UF$_4$.
See Table \ref{table} for further details.

\begin{figure}
\includegraphics[width=3.2in]{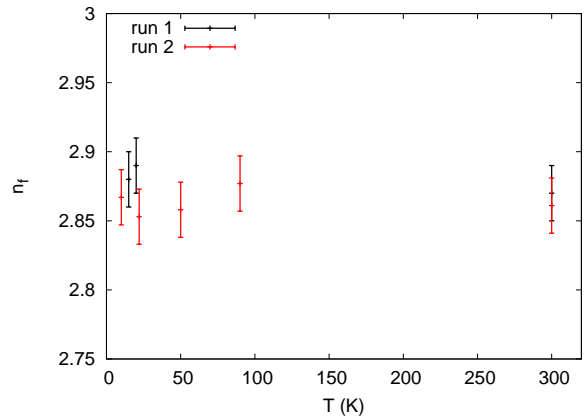}
\caption{Total $5f$ occupancy estimate $n_f$ of \urs\ as a function of 
temperature from both experimental runs. The lowest measured temperature is 
10 K. No measurable temperature dependence was observed. Taking all these data 
together and allowing for some systematic error gives $n_f = 2.87 \pm 0.08$.
}
\label{nf_fig}
\end{figure}

It is crucial to note that the results in Table \ref{table} are for fits to 
data collected only at one temperature. Data collected at other temperatures 
between 10 K and 300 K look very similar, with no clear trend in $n_f$ using 
the Model \#2 constraints (Fig. \ref{nf_fig}). Note that there is no significant
difference between the data collected in the two experimental runs, which were about one
year apart, indicating that if any oxidation on the surface is affecting the 
measurement  (that is, artificially giving too much $f^2$ weight), it is
stable over this time scale. Taken together, we estimate 
$n_f = 2.87 \pm 0.08$,
including possible $\pm 5$\% systematic error in the configuration fractions.

\section{Discussion}
\label{Discussion}

Before discussing the data and their implications for the nature of the $5f$ 
states in \urs, we need to draw the distinction between ``delocalization''
and ``itinerancy'' as they relate to the RXES technique presented here. This
technique is not sensitive to whether given spectroscopic features are 
associated with a band that cuts across the Fermi energy and typically are
described as itinerant. Rather, the technique is sensitive to, as described
in Sec.\ref{Intro}, how well these electrons screen the core hole as a 
consequence of the strength of their Coulomb interaction. This
interaction is substantial for an orbital strongly localized to the vicinity of 
the core hole, such as the $4f$ orbital. A more extended, delocalized orbital, 
like a $d$
orbital or a light-actinide $5f$ orbital, has a weaker interaction, and may not split the 
configurations in Eq. \ref{config_eq}. The observation of a split peak is
therefore a hallmark of a {\em localized} $5f$ orbital (which may still contribute
weakly to the conduction band through hybridization and the Kondo effect), but
the lack of a split peak only indicates a more extended, delocalized orbital,
which may or may not contribute to the Fermi surface. We have therefore 
endeavored to use 
the word ``itinerant'' here only when we are discussing or comparing data to 
experiments indicating a 
Fermi surface or a model with one.


Bearing this distinction in mind, 
there are several useful conclusions to draw by comparing the best fit
parameters from the various materials with those of \urs\ (Table \ref{table}). 
$E_{gi}$, as stated 
above, is arbitrarily set to coincide with the $f^2$ peak position. Unlike in
lanthanide systems (or in the limited number of plutonium systems that have been
measured), $E_{gi}$ can shift to higher energies if an orbital becomes more 
delocalized, and thus fails to screen the core hole as effectively. This change
in screening is the reason that appropriate standard materials are so important,
since the origin of a given feature could be otherwise misinterpreted. Here,
we see that the three relatively localized standard materials have very similar
$E_{gi}$s and the sharpest $\sigma_p$s, with the increased width of UO$_2$
likely due to crystal field splitting of the $d$ states.\cite{Tobin15} Changes
in $E_{if}$ are linear with $E_{gi}$, consistent with no significant 
off-diagonal elements in the transition matrix in this energy range. 

\begin{figure}
\includegraphics[width=3.2in]{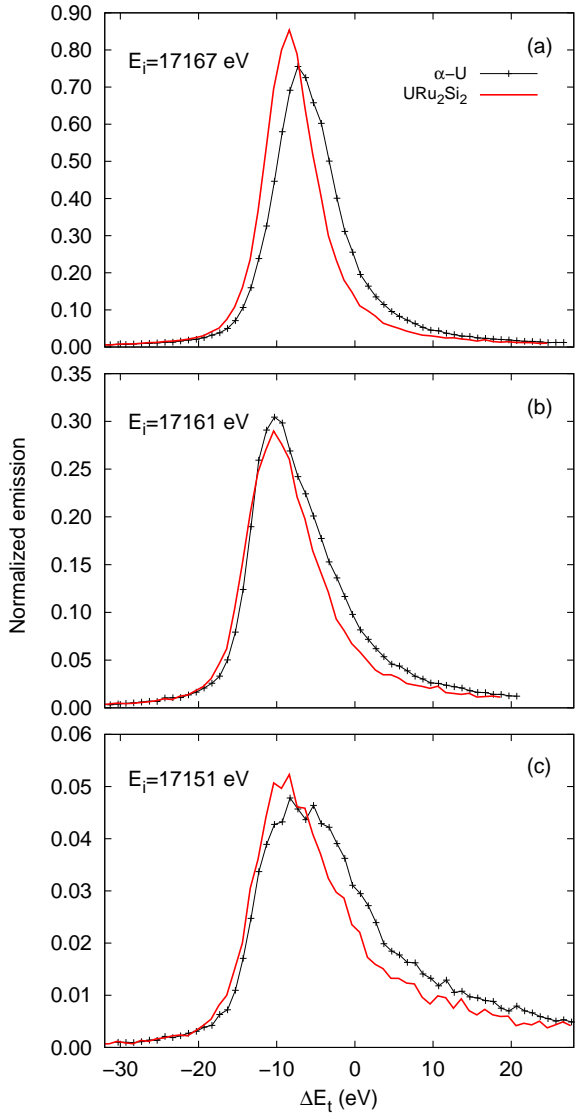}
\caption{Comparison of XES data at 20 K for $\alpha$-U (data from 
Ref.~\onlinecite{Soderlind16}) and \urs\ as a function of $\Delta E_t = E_t-E_{if}$ 
to accentuate possible differences in $5f$ occupancy. Note that these data are 
normalized to the peak XES data with $E_t$ at an incident energy well above 
the absorption edge, in contrast to those in Fig. \ref{stack_fig}.
}
\label{XEScomp_fig}
\end{figure}

It is interesting to compare these data to those from recent results on 
$\alpha$-U,\cite{Soderlind16} which should correspond to an itinerant material. 
The XES spectra of $\alpha$-U shown in Fig. \ref{XEScomp_fig} are clearly
even a little broader than the \urs\ spectra, displaying more indications to
the eye of shoulders and other features indicative of multiple $5f$ 
configurations. The $\alpha$-U data are, in fact, consistent with a local
density of states modeled on two dominant $5f$ configurations, $f^2$ and $f^3$,
corresponding to an $n_f = 2.54 \pm 0.08$. As expected for an itinerant
material, $E_{gi}$ and $E_{if}$ are significantly higher than for the standard 
materials (about 3 eV),
consistent with a more delocalized $5f$ orbital.\cite{Soderlind16}

The most informative fit parameters are those relating to the individual peak 
width 
$\sigma_p$ and the relative configuration fractions. The largest peak widths
here are for $\alpha$-U and \urs, and as such may be indicative of $5f$ orbital
delocalization due to the distribution of possible Coulomb interactions. 
A similar situation is observed for Pu intermetallics,
where the compounds with the lowest linear coefficient to the electronic 
specific heat\cite{Booth12} have the largest peak widths\cite{Booth14} 
(excluding PuO$_{2.06}$ which likely has an enhanced width due to crystal
field splitting of the $d$ states).

The comparisons of the data and fit results between \urs, $\alpha$-U, and the 
standard materials thus strongly favor a large, delocalized $f^3$ component
to the ground state of \urs. In particular, the fit using Model \#2 over 
several temperatures (Fig. \ref{nf_fig}) indicate an $n_f$ of 2.87$\pm$0.08.
Unfortunately, as noted in Sec. \ref{Results}, we do not judge the difference
between fits with Model \#1\ (single $f^3$ configuration) and Model \#2 (a
mixture of $f^2$, $f^3$, and $f^4$ configurations) to be enough to support the
presence of some $f^2$ component due to potential systematic errors, especially 
in the lineshape model. However, a comparison to the $\alpha$-U results supports
this possibility, since the value of $\sigma_p$ for the
Model \#2\ fit is nearly identical to that of 
the fit to $\alpha$-U data where more clearly visible
evidence exists for multiple configurations. 
The fact that multiple excitation peaks, shoulders,
or other visible evidence is not observed in the data in Fig. \ref{XEScomp_fig} 
is because \urs\ is more dominated by a single configuration than $\alpha$-U.
We do not categorically rule out that \urs\ has no $f^2$ component, however.

Taking these results together, we conclude that \urs\ is dominated by a
delocalized $f^3$ configuration and is possibly weakly intermediate
valent. Although it has a similar $n_f$ to strongly localized UCd$_{11}$,
the shift in $E_{gi}$ and the enhanced peak width $\sigma_p$ both indicate a
delocalized $5f$ orbital.

Although $5f$ electron involvement in the conduction band is strongly supported 
by these data, either through a $5f$ band or a Kondo-like mechanism,
it is important to note that the lack of temperature
dependence in $n_f$ is typical of other uranium intermetallics, even in those
with Kondo temperatures between 100 K and 200 K, where one would expect
shifts in $n_f$ at temperatures above 20 K.\cite{BoothPNASnote} 
The lack of temperature dependence here
is consistent with angle-integrated photoemission results.\cite{Yang96}
Like the angle-integrated photoemission experiments, the RXES results 
presented here are sensitive to the average
of all potential $5f$ configurations. We point out that if only a small portion
of the Fermi surface, {\em eg.} the $Z$-point hole pocket, displayed any temperature 
dependence, RXES would not be very sensitive to it.

Although no temperature dependence is observed in these data from room
temperature to 10 K and the measurements are consistent with a partially 
filled $5f$ band, it is instructive to consider the implications in
light of the Anderson model.\cite{Hewson93} As such, this part of the 
discussion is intended to be only qualitative or semi-quantitative in order to 
illustrate the implications of the measurements reported above. 
With a qualitative goal in mind, we can consider these results using
the non-crossing approximation (NCA).\cite{Bickers87} 
Within a simplification 
of this model for a single $f$ electron (which can be taken as a single unpaired
$f$ electron), one can consider:\cite{Cox_thesis}
\begin{equation}
n_f(T) = 1 - \delta n_\text{charge}(T) - \delta n_\text{spin}(T),
\end{equation}
where the Kondo physics affects $\delta n_\text{spin}$, with
\begin{equation}
\delta n_\text{spin}(0) = \frac{\pi T_\text{nca}}{\nu \Gamma} \left(
1+ \frac{\pi T_\text{nca}}{\nu \Gamma} \right)^{-1},
\end{equation}
where $\nu$ is the magnetic degeneracy, 
$\Gamma = \pi \varrho V^2$ is the hybridization strength, 
$\varrho$ is the density of electronic states at the
Fermi energy, $V$ is the hybridization matrix element between $f$ and the 
conduction electrons, and $T_\text{nca}$
is the Kondo temperature as defined in the NCA formalism. 
$\delta n_\text{charge}(T)$ changes very slowly with $T$ where 
$T \ll \epsilon_f$, and so we can consider the low-$T$ value to be constant:
\begin{equation}
\delta n_\text{charge} \equiv \delta n_\text{charge}(0) = 
\frac{\Gamma}{\pi \epsilon_f},
\end{equation}
where $\epsilon_f$ is the absolute energy of the $f$ level with respect to 
$E_F$.
We note that \[\lim_{T \rightarrow \infty} \delta n_\text{spin}(T) = 0,\] 
and 
therefore the high-temperature limit of $n_f$ is $\tilde{n_f} = 
1 - \delta n_\text{charge}$. 

Within this formalism, we expect to observe a total change in $n_f$ of
$\Delta n_f(0) \approx \frac{1}{2} \delta n_\text{spin}(0)$ from $T = 0$ K to about $T \approx T_\text{nca}$. Here, it is important to distinguish between
the coherence temperature $T_\text{coh} \approx 70$ K and the estimate of the
Kondo temperature $T_\text{K} \approx 370$ K.\cite{Schoenes87} For this
rough discussion, taking $T_\text{K} \sim T_\text{nca} \sim 300$ K and 
including no more than $\Delta n_f(0) \lesssim 0.05$ {\em limits} the ratio of 
$T_\text{nca}/\Gamma$ to about 0.3. We have confirmed this limit with a more 
detailed NCA calculation.\cite{Lawrence01} For smaller $T_\text{nca}$, this limit 
would be even 
more restrictive. Therefore, the main conclusion of this qualitative discussion 
is simply that charge fluctuations dominate the interpretation of the RXES 
data.  We stress that this estimate is in the single-impurity regime above the
coherence temperature, and definitely above the HO transition, below which
Hall effect measurements indicate a very small carrier concentration of
only $\sim$0.05 holes per formula unit within the usual 1-band approximation. 
Such a low carrier concentration would not provide enough conduction holes to 
quench any unpaired
$5f$ electron spins in any Kondo effect,\cite{Burdin00} a situation
particularly important when one moves away from a single-impurity
model and toward a lattice model.\cite{Bauer04}

The picture that emerges from these data is therefore one that is dominated
by conventional charge fluctuations such as one would expect from itinerant 
$5f$ electrons, with very little if any temperature dependence indicating 
strong hybridization between the $5f$ orbital and the conduction band. On the 
other hand, a contribution from an $f^2$ configuration is consistent with
the RXES data. It remains possible that the delocalized $f^3$-like contributions
originate from the majority portion of the Fermi surface, while a minority
portion, such as the $Z$-point hole pocket, are the source of more localized
$f^2$-like behavior. If this is the case, any temperature dependence in the
U \liii-edge absorption of this minority portion would be obscured by the
majority $f^3$-like portion. In other words, these data are
easily rectified with the itinerant $f^3$ band theories, 
but could still allow for a localized $f^2$ theory
if that theory only applied to a small portion 
of the Fermi surface or some other minority portion of the electronic structure. 
This dichotomy is therefore suggestive of a 2-fluid like interpretation, and, 
in fact, 2-fluid theories appear in many explanations of various actinide 
phenomena,\cite{Schoenes96,Zwicknagl03, Steglich97,Cox96,Nakatsuji04,Yang12} 
including of \urs.\cite{Okuno98,Shirer12}

\section{Conclusion}

RXES measurements at the U \liii\ edge and the U $L_{\alpha1}$ emission
indicate that \urs\ has a delocalized $5f$ orbital with a mean occupancy 
$n_f = 2.87 \pm 0.08$.  The conclusion
of a delocalized orbital is derived from the line shape of the RXES signal,
by the shift in the \liii\ threshold energy, and the lack of temperature 
dependence.  These results are consistent with EELS and photoemission
experiments (see Sec. \ref{Intro}). These results are not consistent with
theoretical models that require a localized $f^2$ state to generate crystal
field splitting in the $5f$ manifold, unless this state could be in the
minority compared to a majority $f^3$ band. 

\section*{Acknowledgments}

CHB acknowledges several useful conversations with Jason Jeffries, 
Nicholas Butch, Andrew Wray, Jonathon Denlinger, and Jon Lawrence. 
JGT gratefully acknowledges support during FY 2014 from the Lawrence Livermore 
National Laboratory (LLNL) PRT program for his sabbatical at Lawrence Berkeley 
National Laboratory (LBNL).  Work at LBNL was supported by the Director,
Office of Science, Office of Basic Energy Sciences (OBES), of the U.S.
Department of Energy (DOE) under contract DE-AC02-05CH11231.  
LLNL is operated
by Lawrence Livermore National Security, LLC, for
the U.S. Department of Energy, National Nuclear Security
Administration, under Contract DE-AC52-07NA27344.
RXES data were collected at the Stanford Synchrotron Radiation Lightsource,
a national user facility operated by Stanford University on behalf of the DOE,
OBES.  Work at Los Alamos National Laboratory was performed under the auspices
of the U.S.  DOE, OBES, Division of Materials Sciences and Engineering.

\bibliography{/home/hahn/chbooth/papers/bib/bibli}

\begin{thebibliography}{66}
\bibitem[{Boo()}]{BoothPNASnote}
\bibinfo{note}{Although not explicitly mentioned in Ref. \onlinecite{Booth12},
  all data reported therein were collected at temperatures of 20 or 30 K and at
  room temperature.}
\end{thebibliography}

\end{document}